\documentclass[12pt]{article}
\usepackage{graphicx,amssymb}

\newcommand{\e}{{\rm e}}

\title{Microscopic model of phase transition in the crystals of
DMAAlS and DMAGaS type}
\author{I.V.Stasyuk, O.V.Velychko \\ [2ex]
Institute for Condensed Matter Physics\\
of the National Academy of Sciences of Ukraine,\\
1 Svientsitskii Str., 79011 Lviv, Ukraine
}
\date{}

\begin{document}

\maketitle

\centerline{\parbox{12cm}{\small
The four-state model is proposed for description of phase transition in
ferroelectric crystals of DMAGaS and DMAAlS type.  Thermodynamical functions
of the model are obtained in the mean field approximation.  The phase
transition between paraelectric and ferroelectric phases is investigated.
It is established that order of
the phase transitions depends on relations between model parameters.
{
}
}}

\section{Introduction}

A peculiar feature of isomorphous crystals with ferroelectric properties
\linebreak
(CH$_3$)$_2$NH$_2$Al(SO$_4$)$_2$ $\cdot$ 6H$_2$O (DMAAlS) and
(CH$_3$)$_2$NH$_2$Ga(SO$_4$)$_2$ $\cdot$ 6H$_2$O (DMAGaS) is the
possibility of existence in three different phases at change of
temperature: at room temperature crystal is paraelectric, at lowering of
temperature it sequentially becomes ferroelectric and further can be in
antiferroelectric state \cite{PJCh93,PJCh94,PJCh95,ZNat98}.

\begin{figure}
\centerline{\includegraphics[width=0.6\textwidth]{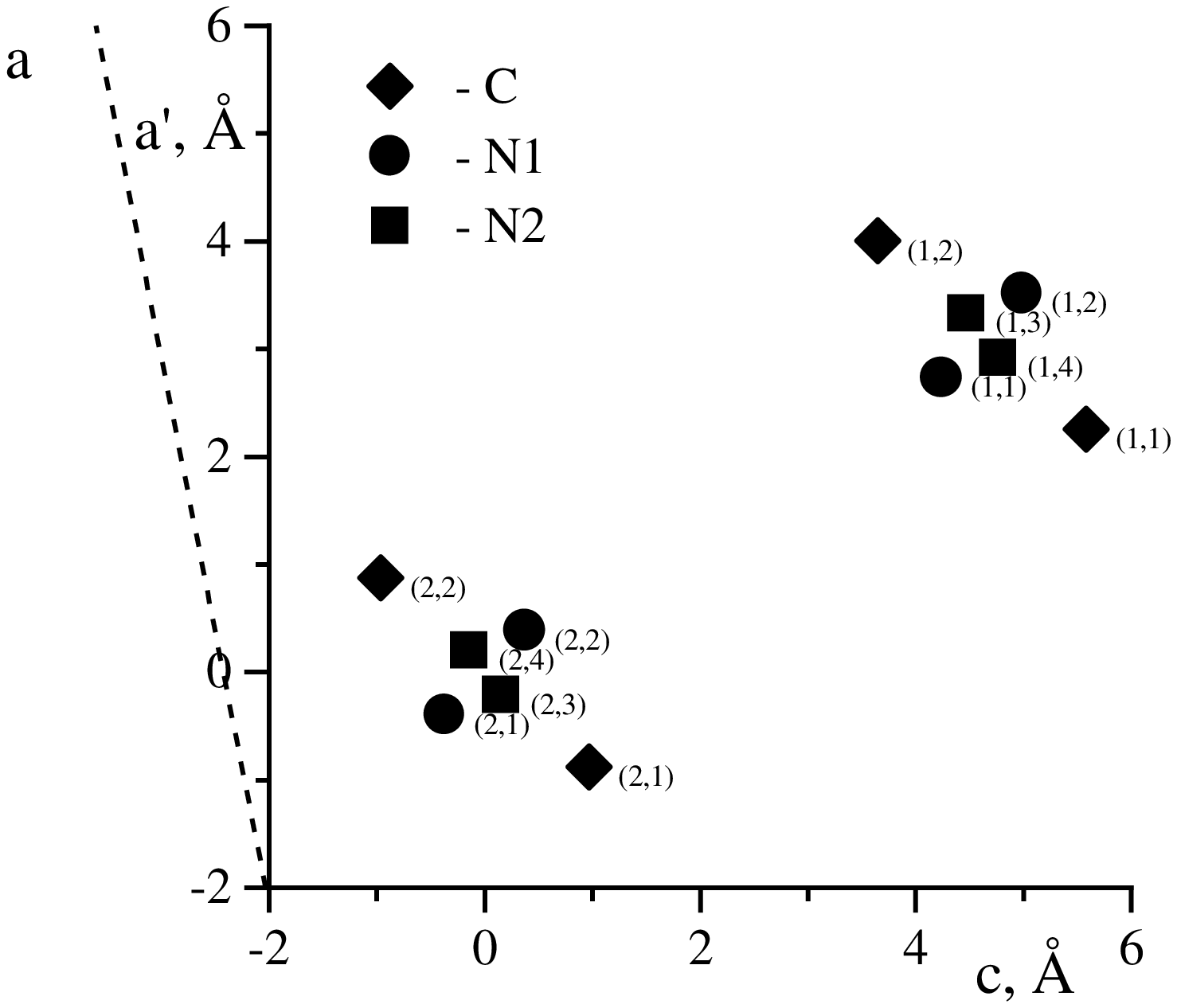}}
\caption{%
Projection of N and C atoms in the dimethylammonium groups of
DMAGaS crystal at ambient temperature \protect\cite{PJCh93} onto the XZ
plane (position indices are indicated in parenthesis).
The Y coordinates of atoms are as follows:
Y$_{N(1,1)}$=--0.1469~\AA, Y$_{N(1,2)}$=0.1469~\AA,
Y$_{N(1,3)}$=--0.6736~\AA, Y$_{N(1,4)}$=0.6736~\AA,
Y$_{N(2,1)}$=5.5099~\AA, Y$_{N(2,2)}$=5.2161~\AA,
Y$_{N(2,3)}$=4.6894~\AA, Y$_{N(2,4)}$=6.0366~\AA,
Y$_{C(1,1)}$= --0.2606~\AA, Y$_{C(1,2)}$=0.2606~\AA,
Y$_{C(2,1)}$=5.6236~\AA, Y$_{C(2,2)}$=5.1024~\AA.}
\label{xzprj}
\vspace{5ex}

\centerline{\includegraphics[width=0.6\textwidth]{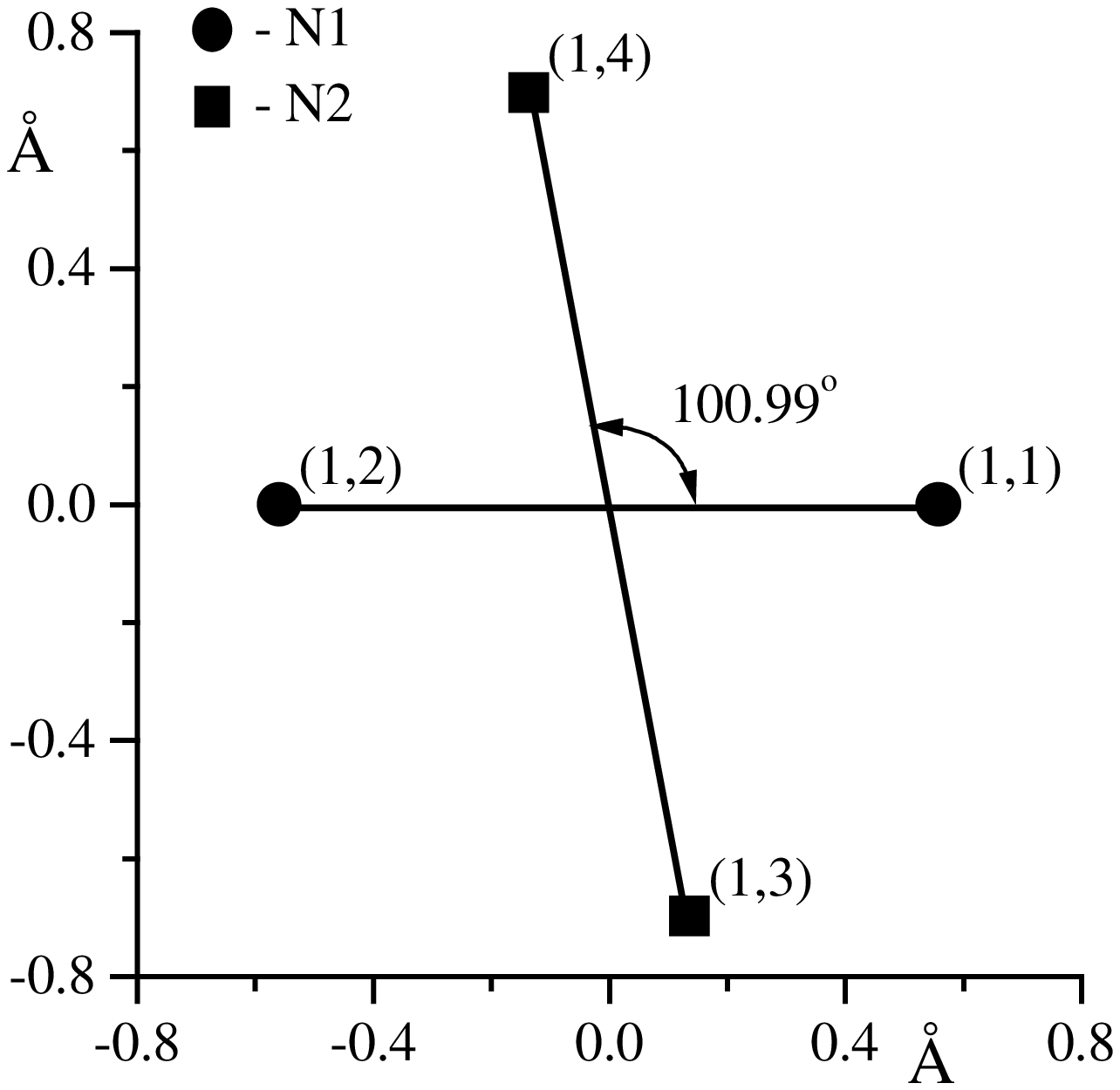}}
\caption{%
Projection of N atoms onto the plane
perpendicular to the C--C$'$ axis of the dimethylammonium group.}
\label{cross}
\end{figure}

It has been measured $T_{c1}=136$~K and $T_{c2}=113$~K
\cite{ZNat98} (or 122~K and 114~K respectively \cite{PJCh95}) for
DMAGaS crystal, but it has been only found $T_{c1}=150$~K
\cite{PJCh93,PJCh94,PJCh95} for DMAAlS. The phase transition between the
ferroelectric and antiferroelectric phases is of the first order
\cite{PJCh95,ZNat98}.
The phase transition paraelectric -- ferroelectric is of the first
order close to the second one in DMAGaS and of the second order in DMAAlS.
Crystallographic analysis shows that in all three phases crystal
belongs to monoclinic space groups: high-temperature paraelectric phase
has P2$_1$/n space group \cite{PJCh93,PJCh94}, ferroelectric and
antiferroelectric ones have Pn \cite{PJCh94,PJCh95} and P2$_1$
\cite{PJCh95} groups respectively. It should be mentioned that
low-symmetry space groups are subgroups of the high-symmetry group
obtained by the loss of rotation axis and mirror
plane, respectively (point symmetry group 2/m changes to m or 2).

Data of investigation of various physical properties of the crystals were
presented in a range of publications. First of all it should be mentioned
dielectric, pyroelectric and dilatometric measurements
\cite{ZNat98,r5,r6,r7,r8,r9,r11,r12,a1}, ultrasonic and NMR investigations
\cite{r10,r14,a2}, studies of relaxational dynamics and lattice dynamics
\cite{a3,a4} by means of radiospectroscopy and Raman spectroscopy methods.
Such
issues as dielectric anomalies in the vicinity of phase transition points
(mainly near $T_{c1}$), clarification of the order of phase transition,
dynamics of ionic groups and protons on hydrogen bonds in the wide range of
temperatures \cite{a3,a4}, influence of external hydrostatic pressure
\cite{r16,a5,a6} were considered. Temperatures of phase transitions were
justified; according to the latest data $T_{c1}=136$~K, $T_{c2}=117$~K
(DMAGaS) and $T_{c1}=155$~K (DMAAlS).

Crystals DMAAlS and DMAGaS belong to the type of ferroelectrics which have
some structural elements which reorientation leads to polarisation of
crystal.  In this case the element, which can be reoriented, is the
dimethylammonium cation (more strictly NH$_2$ group). Crystals contain two
symmetrically nonequivalent DMA groups per unit cell (Fig.~\ref{xzprj}).

DMA can
occupy four equilibrium positions which are related in pairs by inversion
centre forming a slightly deformated cross (Fig.~\ref{cross}). It is very
likely that this asymmetry (and corresponding difference of energies of
interaction between groups in various positions) is responsible for so
complicated behaviour of the crystal.
In the paraelectric phase the site in one pair is occupied with
probability 40\% and in another with probability 10\% at 300~K.

Possible role of reorientations of DMA groups in phase transitions of the
considered family of crystals was pointed out in a range of articles
\cite{PJCh95,r6,r14,r13}. This assumption is directly confirmed by the data of
structural investigations \cite{r14,r13}. Performed recently NMR
measurements \cite{r14} have proved that reorientations and orderings of DMA
groups can be considered as an origin of ferroelectrics
(F) and antiferroelectrics
(AF) phase transitions while other changes in a
lattice structure (rotations of SO$_4$ groups, freezing of protons in certain
positions on hydrogen bonds) are just accompanying phenomena.

So for theoretical description of phase transitions to ferro- and
antiferroelectric states in DMAAlS and DMAGaS crystals and development of
appropriative microscopic model one should take into account reorientation
of DMA groups on and their influence on the state of crystal. Such an
approach is
a main goal of the present paper. Our task is  to formulate
this type model, to introduce order parameters responsible for appearance of
low-symmetry phases, to calculate thermodynamical potential proceeding from
the model and to investigate thermodynamically stable states. The transition
from paraelectric to ferroelectric state and the dependence of the order
of this transition on the
system parameter values will be studied in more detail.

\section{Hamiltonian of the model}

In both high- and low-temperature phases ionic groups belong to two
different sublattices (an elementary cell consist of two translationally
nonequivalent groups). We shall characterize orientations of the groups by
the spatial localization of nitrogen atoms belonging to the
groups. Four possible positions of $N$ atom are described by the Hubbard
projection operators $X_{nk}^{pp}$, where $n$ is the number of a lattice site
(an elementary cell), $\alpha=1,2$ is the sublattice index, $p=1,\ldots,4$
is the position number ($X_{n\alpha}^{pp}=0$ or 1; $\sum_{p}
X_{n\alpha}^{pp}=1$). Energies of different positions are pairwise
equivalent: $\varepsilon_{1}=\varepsilon_{2}$, $\varepsilon_{3}=\varepsilon_{4}$ (see
notations in Figs.~\ref{xzprj} and~\ref{cross}).

The model Hamiltonian can be written down as follows:
\begin{equation}
H =\sum_{n\alpha{}p} \varepsilon_p X_{n\alpha}^{pp}
   -\frac12 \sum_{nn'}\sum_{\alpha\alpha'\atop{}pp'}
   J_{\alpha\alpha'}^{pp'}(nn')
   X_{n\alpha}^{pp} X_{n'\alpha'}^{p'p'},
\label{1.1}
\end{equation}
where the first and the second terms describe energy of separate complexes
and pair interaction $J_{\alpha\alpha'}^{pp'}(nn')$ depending on their
orientational states, respectively.

Further consideration will be performed in the mean field approximation. By
substitution $XX\to 2\langle X\rangle X - \langle X\rangle \langle X\rangle$
in (\ref{1.1}) one can obtain
\begin{equation}
H_{\rm MF}=NH_{\rm c} + \sum_{n\alpha p} [\varepsilon_{p} - f_{\alpha p}]
X_{n\alpha}^{pp},
\label{1.2}
\end{equation}
where the mean field is introduced
\begin{equation}
f_{\alpha p} = \sum_{n'\alpha 'p'} J_{\alpha\alpha '}^{pp'} (nn') \langle
X_{n'\alpha '}^{p'p'}\rangle
\label{1.3}
\end{equation}
and
\begin{equation}
H_{\rm c}=\frac{1}{2} \sum_{n'} \sum_{{{\alpha\alpha '}\atop{pp'}}}
J_{\alpha\alpha '}^{pp'} (nn') \langle X_{n\alpha}^{pp} \rangle \langle
X_{n'\alpha '}^{pp'} \rangle
\label{1.4}
\end{equation}

Averages $\langle X_{n\alpha}^{pp}\rangle$ do not depend on cell number
($\langle X_{n\alpha}^{pp}\rangle\equiv \langle X_{\alpha}^{pp}\rangle$) due
to translational symmetry which does not change at the transition to low-symmetry
phases (e.g. lattice period is not doubled in antiferroelectric phase). The
averages have meaning of occupation numbers of corresponding orientational states
and can be grouped in linear combinations
\begin{equation}
\xi_{\mu} = \sum_{\alpha p} U_{\mu,\alpha p} \langle X_{\alpha}^{pp}\rangle,
\label{1.5}
\end{equation}
which transform according to irreducible representations of point symmetry group
$2/m$ of high-temperature phase. Coefficients of transformation (\ref{1.5})
are presented in Tab.~\ref{tab1}. As one can see all possible irreducible
representations (A$_{g}$, B$_{g}$, B$_{u}$, A$_{u}$) are realized, two linear
combinations of a type of (\ref{1.5}) belong to each of them.

Combinations $\xi_{\mu}$, which are transformed according to $B_{u}$ and
$A_{u}$ representations, are connected with differences of occupations of
($\alpha,1$), ($\alpha,2$) and ($\alpha,3$), ($\alpha,4$) positions and can
play a role of order parameters for phase transitions to ferroelectric
($B_{u}$) and antiferroelectric ($A_{u}$) phases.

As one can see in Tab.~\ref{tab1} the first of parameters belonging to the $B_{u}$
representation ($y_{+}$) describes ferroelectric ordering along the
ferroelectric axis (the OX' axis in the crystallographic plane (ac)) while
the second one ($u_{-}$) corresponds to antiferroelectric ordering along OY
(the crystallographic axis b). For the $A_{u}$ representation the ordering
is inverse: the parameter $y_{-}$ describes antiferroelectric ordering along
OX' and the parameter $u_{+}$ defines ferroelectric ordering along OY. Thus,
the states of ferro- and antiferrophase observed in DMAAlS and DMAGaS
crystals in reality are mixed.

Using (\ref{1.3}) and (\ref{1.5}) one can write down
\begin{equation}
f_{\alpha p} = \sum_{\mu\mu '} (U^{-1})_{\alpha p,\mu} \tilde{j}_{\mu\mu '}
\xi_{\mu '},
\label{1.6}
\end{equation}
where
\begin{eqnarray}
\tilde{j}_{\mu\mu '} &=& \sum_{\alpha p}\sum_{\alpha ' p'} U_{\mu,\alpha p}
j_{\alpha\alpha'}^{pp'} (U^{-1})_{\alpha ' p',\mu '}\:, \nonumber\\
j_{\alpha\alpha '}^{pp'} &=& \sum_{n'} J_{\alpha\alpha '}^{pp'} (nn');
\label{1.7}
\end{eqnarray}
and
\begin{equation}
H_{\rm c}  = \frac{1}{2} \sum_{\mu\mu '} \tilde{j}_{\mu\mu '} \xi_{\mu}
\xi_{\mu '}.
\label{1.8}
\end{equation}

\section{Free energy and equations for order parameters}

Proceeding from expressions (\ref{1.2}), (\ref{1.6}) and (\ref{1.8}) one can
easy derive main thermodynamical functions of the model in the mean field
approximation and construct a set of equations for selfconsistency
parameters $\xi_{\mu}$.

Corresponding partition function is
\begin{equation}
Z = \e^{-\beta{}NH_c}
\left[\sum_{p=1}^4 \e^{-\beta(\varepsilon_p-f_{1p})}\right]^N
\left[\sum_{p=1}^4 \e^{-\beta(\varepsilon_p-f_{2p})}\right]^N
\label{1.10}
\end{equation}
or
\begin{equation}
Z = \e^{-\beta{}NH_c}\left[2Z_1\right]^N\left[2Z_2\right]^N,
\label{1.11}
\end{equation}
where
\begin{eqnarray}
Z_1 &=&
Q_x^{+}  \cosh\beta(k_{y+}+k_{y-}) +
Q_z^{+}  \cosh\beta(k_{u-}+k_{u+}),
\label{1.12}\\
\nonumber
Z_2 &=&
Q_x^{-}   \cosh\beta(k_{y+}-k_{y-}) +
Q_z^{-}   \cosh\beta(k_{u-}-k_{u+}),
\end{eqnarray}
and
\begin{eqnarray}
&&
Q_x^{\pm} = \e^{-\beta\varepsilon_1} \exp[-\beta(k_{x+} \pm k_{x-})],\quad
Q_z^{\pm} = \e^{-\beta\varepsilon_3} \exp[-\beta(k_{z+} \pm k_{z-})],\\
\nonumber
&&
k_{x+} = (A_1 x_+ + C_1 z_+)/2,\quad
k_{x-} = (A_2 x_- + C_2 z_-)/2,\\
\nonumber
&&
k_{z+} = (C_1 x_+ + B_1 z_+)/2,\quad
k_{z-} = (C_2 x_- + B_2 z_-)/2,\\
&&
k_{y+} = (A_3 y_+ + C_3 u_-)/2,\quad
k_{y-} = (A_4 y_- + C_4 u_+)/2,\label{1.13}\\
\nonumber
&&
k_{u-} = (C_3 y_+ + B_3 u_-)/2,\quad
k_{u+} = (C_4 y_- + B_4 u_+)/2.
\end{eqnarray}
Here such notations are introduced
\begin{eqnarray}
\nonumber
&& A_1 = a + b + g + h,\;
   B_1 = e + f + l + m,\;
   C_1 = c + d + j + k,\\
&& A_2 = a + b - g - h,\;
   B_2 = e + f - l - m,\;
   C_2 = c + d - j - k,\\
\nonumber
&& A_3 = a - b + g - h,\;
   B_3 = e - f - l + m,\;
   C_3 = c - d - j + k,\\
\nonumber
&& A_4 = a - b - g + h,\;
   B_4 = e - f + l - m,\;
   C_4 = c - d + j - k,\\
\nonumber
&&
a = j_{11}^{11}(0),\; b = j_{11}^{12}(0),\;
c = j_{11}^{13}(0),\; d = j_{11}^{14}(0),\\
&&
e = j_{11}^{33}(0),\; f = j_{11}^{34}(0),\;
g = j_{12}^{11}(0),\; h = j_{12}^{12}(0),\label{a1.7}\\
\nonumber
&&
j = j_{12}^{13}(0),\; k = j_{12}^{14}(0),\;
l = j_{12}^{33}(0),\; m = j_{12}^{34}(0).
\end{eqnarray}

Free energy per site $F$ is
\begin{equation}
F = H_c-\Theta\ln(2Z_1)-\Theta\ln(2Z_2).
\label{1.14}
\end{equation}

 By direct averaging
or with the use of free energy~(\ref{1.14}) one can obtain the set of
equations for the parameters $\xi_{\mu}$:
\begin{eqnarray}
\nonumber
x_{\pm} &=& [Z_1^{-1} Q_x^{+} \cosh\beta(k_{y+}+k_{y-}) \pm
              Z_2^{-1} Q_x^{-} \cosh\beta(k_{y+}-k_{y-})]/2,
\\
z_{\pm} &=& [Z_1^{-1} Q_z^{+} \cosh\beta(k_{u+}+k_{u-}) \pm
              Z_2^{-1} Q_z^{-} \cosh\beta(k_{u+}-k_{u-})]/2,
\label{1.15}\\
\nonumber
y_{\pm} &=& [Z_1^{-1} Q_x^{+} \sinh\beta(k_{y+}+k_{y-}) \pm
              Z_2^{-1} Q_x^{-} \sinh\beta(k_{y+}-k_{y-})]/2,
\\
\nonumber
u_{\pm} &=& [Z_1^{-1} Q_z^{+} \sinh\beta(k_{u+}+k_{u-}) \pm
              Z_2^{-1} Q_z^{-} \sinh\beta(k_{u+}-k_{u-})]/2.
\end{eqnarray}

\begin{table}
\begin{center}
\renewcommand{\arraystretch}{0}
\begin{tabular}{|c|c||c|c|c|c|c|c|c|c|}
\hline
&&(1,1)&(1,2)&(1,3)&(1,4)&(2,1)&(2,2)&(2,3)&(2,4)\strut\\
\hline
\rule{0pt}{2pt}&&&&&&&&&\\
\hline
\raisebox{-1.7ex}[0pt][0pt]{A$_g$}
      & $x_+$& 1/2&  1/2&  0&  0&  1/2&  1/2&  0&  0\strut\\
\cline{2-10}
      & $z_+$& 0&  0&  1/2&  1/2&  0&  0&  1/2&  1/2\strut\\
\hline
\raisebox{-1.7ex}[0pt][0pt]{B$_g$}
      & $x_-$& 1/2&  1/2&  0&  0& -1/2& -1/2&  0&  0\strut\\
\cline{2-10}
      & $z_-$& 0&  0&  1/2&  1/2&  0&  0& -1/2& -1/2\strut\\
\hline
\raisebox{-1.7ex}[0pt][0pt]{B$_u$}
      & $y_+$& 1/2& -1/2&  0&  0&  1/2& -1/2&  0&  0\strut\\
\cline{2-10}
      & $u_-$& 0&  0&  1/2& -1/2&  0&  0& -1/2&  1/2\strut\\
\hline
\raisebox{-1.7ex}[0pt][0pt]{A$_u$}
      & $y_-$& 1/2& -1/2&  0&  0& -1/2&  1/2&  0&  0\strut\\
\cline{2-10}
      & $u_+$& 0&  0&  1/2& -1/2&  0&  0&  1/2& -1/2\strut\\
\hline
\end{tabular}
\renewcommand{\arraystretch}{1}
\end{center}
\caption{Coefficients of symmetrized occupancies of orientational states
which correspond to irreducible representations of the point symmetry
group 2/m.}
\label{tab1}
\end{table}

\section{Phase transition between paraelectric and ferroelectric
phases}

Let us consider phase transition (PT) from paraelectric
phase to the ferroelectric one which corresponds to irreducible
representation $B_u$ with the two-component order parameter (OP). In this
case averages $x_-$, $z_-$, $y_-$ and $u_+$ are equal to zero, averages
$y_+$ and $u_-$ are the components
of OP and $Z_1=Z_2$. Then the expression for free energy (\ref{1.14})
simplifies
\begin{eqnarray}
F &=& H_c-2\Theta\ln(2Z_1), \label{2.1}\\
\nonumber
H_c &=& \frac12
\left(
A_1 N_1^2 + 2 C_1 N_1 N_2 + B_1 N_2^2 +
A_3 \xi^2 + 2 C_3 \xi \eta + B_3 \eta^2
\right),\\
\nonumber
Z_1 &=&
\e^{-\beta\varepsilon_1}
\exp\left[-\frac12\beta(A_1 N_1 + C_1 N_2)\right]
\cosh\left[\frac12\beta(A_3 \xi + C_3 \eta)\right]\\
\nonumber
&&
{}+
\e^{-\beta\varepsilon_3}
\exp\left[-\frac12\beta(C_1 N_1 + B_1 N_2)\right]
\cosh\left[\frac12\beta(C_3 \xi + B_3 \eta)\right].
\end{eqnarray}
Here
\begin{equation}
N_1 = x_+, N_2 = z_+, \xi = y_+, \eta = u_-.
\label{2.2}
\end{equation}
It is convenient to separate the term in the expression for the free energy
which depends on $\xi$ and $\eta$
\begin{eqnarray}
F &=& F_0 + F_1(\xi,\eta), \label{2.3}\\
\nonumber
F_0 &=& \frac12
\left(
A_1 N_1^2 + 2 C_1 N_1 N_2 + B_1 N_2^2
\right)
-2\Theta \ln 2 -2\Theta \ln (a+b),\\
F_1(\xi,\eta) &=&
\left(
A_3 \xi^2 + 2 C_3 \xi \eta + B_3 \eta^2
\right)
\label{2.4}
\\
\nonumber
&&
{-}2\Theta \ln
\left\{
n_1 \cosh\left[\frac12\beta(A_3 \xi {+} C_3 \eta)\right]
+
n_2 \cosh\left[\frac12\beta(C_3 \xi {+} B_3 \eta)\right]
\right\},
\end{eqnarray}
where the following symbols are introduced
\begin{eqnarray}
\nonumber
&&
a =
\e^{-\beta\varepsilon_1}
\exp\left[-\frac12\beta(A_1 N_1 + C_1 N_2)\right]
,\;
b =
\e^{-\beta\varepsilon_3}
\exp\left[-\frac12\beta(C_1 N_1 + B_1 N_2)\right]
,\\
&&
n_1 = a/(a+b),\; n_2 = b/(a+b),\; n_1 = 1-n_2.
\label{add1}
\end{eqnarray}

Expressions $n_1$ and $n_2$ describe an effective occupancy of equilibrium
positions 1,2 and 3,4, respectively, in disordered state. One can assume they
have a weak dependence on temperature in the region of the
paraelectric-ferroelectric phase transition.
In the
framework of this approach $n_1$ and $n_2$ could be considered as model
parameters.
The main contribution to the change
of free energy is given by the term $F_1(\xi,\eta)$.   Then
the averages $\xi$ and $\eta$ are given by a set of equations
\begin{equation} \begin{array}{rcl}
\xi &=& \displaystyle
\frac%
{(1-n)\sinh[\tilde{\beta}(\xi+C\eta)]}%
{(1-n)\cosh[\tilde{\beta}(\xi+C\eta)]+n\cosh[\tilde{\beta}(C\xi+B\eta)]}\\
\rule{0pt}{6ex}
\eta &=& \displaystyle
\frac%
{n\sinh[\tilde{\beta}(C\xi+B\eta)]}%
{(1-n)\cosh[\tilde{\beta}(\xi+C\eta)]+n\cosh[\tilde{\beta}(C\xi+B\eta)]}\\
\end{array}
,
\label{2.5}
\end{equation}
where following symbols are used for normalized parameters
$$
B=B_3/A_3,\; C=C_3/A_3,\; \tilde{\beta}=\beta A_3/2,\; n=n_2\,.
$$
The set of equations (\ref{2.5}) is invariant with respect to the inversion
operation $\xi,\eta \to -\xi,-\eta$ and has, as usually, a trivial
paraelectric solution $\xi=\eta=0$. It can be
 established that at $\Theta=0$ two
another solutions ($\xi=0, \eta=1$ at $C<B$ and $\xi=1, \eta=0$ at $C<1$)
exist.

Landau expansion of the free energy is a convenient tool for investigation
of PTs of the second order (PT2) and of the first order (PT1) which are
close to second order. But in this case the use of Landau expansion meets
with problems due to a negative sign of the coefficient at the sixth order
of OP for the some range of parameter values. For this reason Landau
expansion plays an auxiliary role in our investigations. Besides that we
perform all calculations of $\xi$ and $\eta$ averages, free energy and the
temperature of PT numerically using the expression (\ref{2.3}) and the set
(\ref{2.5}).

Landau expansion up to fourth order was used for calculation of
temperature of possible PT2 and boundaries of the regions where PT1 can
occur. In this case OP is two-component, hence for diagonalization of the
quadratic form, which looks like
\begin{eqnarray}
&&
F_1(\xi,\eta) = U\xi^2+V\eta^2+2S\xi\eta, \label{2.6}\\
&&
U=\frac12\left[1-\left(1+(C^2-1)n\right)\frac{\tilde{\beta}}{2}\right],
V=\frac12\left[B-\left(C^2+(B^2-C^2)n\right)\frac{\tilde{\beta}}{2}\right],\\
\nonumber
&&
S=\frac{C}{2}\left[1-\left(1+(B-1)n\right)\frac{\tilde{\beta}}{2}\right],
\label{2.7}
\end{eqnarray}
one should make the following transformation
\begin{eqnarray}
&&
\left(
\begin{array}{c}
r_1 \\ r_2
\end{array}
\right)
= \hat{T}
\left(
\begin{array}{c}
\xi \\ \eta
\end{array}
\right),
\quad
\hat{T} =
\left(
\begin{array}{cc}
\cos\varphi & -\sin\varphi \\
\sin\varphi & \cos\varphi
\end{array}
\right),
\label{2.8}\\
&&
\cos 2\varphi = (V-U)/\sqrt{(V-U)^2+4S^2},\;
\sin 2\varphi = 2S/\sqrt{(V-U)^2+4S^2}.
\label{2.9}
\end{eqnarray}
Now the expression (\ref{2.6}) is as follows
\begin{eqnarray}
&&
F_2(r_1,r_2) =
\Lambda_1 r_1^2 + \Lambda_2 r_2^2, \label{2.10}\\
\nonumber
&&
\Lambda_1 = \frac12\left[(V+U)-\sqrt{(V-U)^2+4S^2}\right],\;
\Lambda_2 = \frac12\left[(V+U)+\sqrt{(V-U)^2+4S^2}\right].
\label{2.11}
\end{eqnarray}
Coefficients $\Lambda_1$ and $\Lambda_2$ are equal to zero at temperatures
\begin{equation}
\tilde{\Theta}_{1,2} = \left[(B-C^2)(1-n)n\right]
\Bigg/
\left[
\left(1-(1-B)n\right)\mp
\sqrt{\left(1-(1+B)n\right)^2+4C^2(1-n)n}\,
\right].
\end{equation}
One can obtain conditions on parameters $A_3$, $B_3$ and $C_3$
\begin{equation}
A_3>0,\; B_3>0,\;  A_3 B_3>C_3; \quad B>0,\; B>C^2,
\label{2.13}
\end{equation}
taking into account that at $\Theta=\infty$ ($\beta=0$) the system is
paraelectric (coefficients $\Lambda_1$ and $\Lambda_2$ are positive). If
the conditions (\ref{2.13}) are satisfied, the temperature
$\tilde{\Theta}_1$ is always higher than the temperature
$\tilde{\Theta}_2$. Hence it determines the temperature of possible PT2
and variable $r_1$ in the expression (\ref{2.10}) becomes the OP.

The region of parameter values, where PT1 takes place, can be established
by the criterion of a negative fourth order term proportional to $r_{1}^{4}$
of Landau expansion. This term can be obtained from the expression
\begin{eqnarray}
\nonumber
F_1^{(4)}(\xi,\eta) &=&
-\frac{\tilde{\beta}}{32}\bigg\{
\left( 2 k_1 + 2 k_2 C^4 - (1-n) n C^2 \right) \xi^4\\
\nonumber
&&
\quad{}+
\left( 2 k_1 C^4 + 2 k_2 B^4 - (1-n) n B^2 C^2 \right) \eta^4 \\
&&
\quad{}+
2 C \left( 4 k_1 + 4 k_2 B C^2 - (1-n) n (B+C^2) \right) \xi^3\eta\\
\nonumber
&&
\quad{}+
2 C \left( 4 k_1 C^2 + 4 k_2 B^3 - (1-n) n B(B+C^2) \right) \xi\eta^3
\label{2.14}\\
\nonumber
&&
\quad{}+
\left( 12(k_1 + k_2 B^2)C^2 - (1-n) n C^2 (B^2 + 4 B C^2 + C^4) \right)
\xi^2\eta^2
\bigg\},\\
\nonumber
&&
k_1 = (1-n)/12-(1-n)^2/4,\; k_2 = n/12-n^2/4,
\end{eqnarray}
using the transformation (\ref{2.8}).

\begin{figure}
\begin{center}
\begin{tabular}{p{0.45\textwidth}p{0.45\textwidth}}
\hfil
\includegraphics[width=0.35\textwidth]{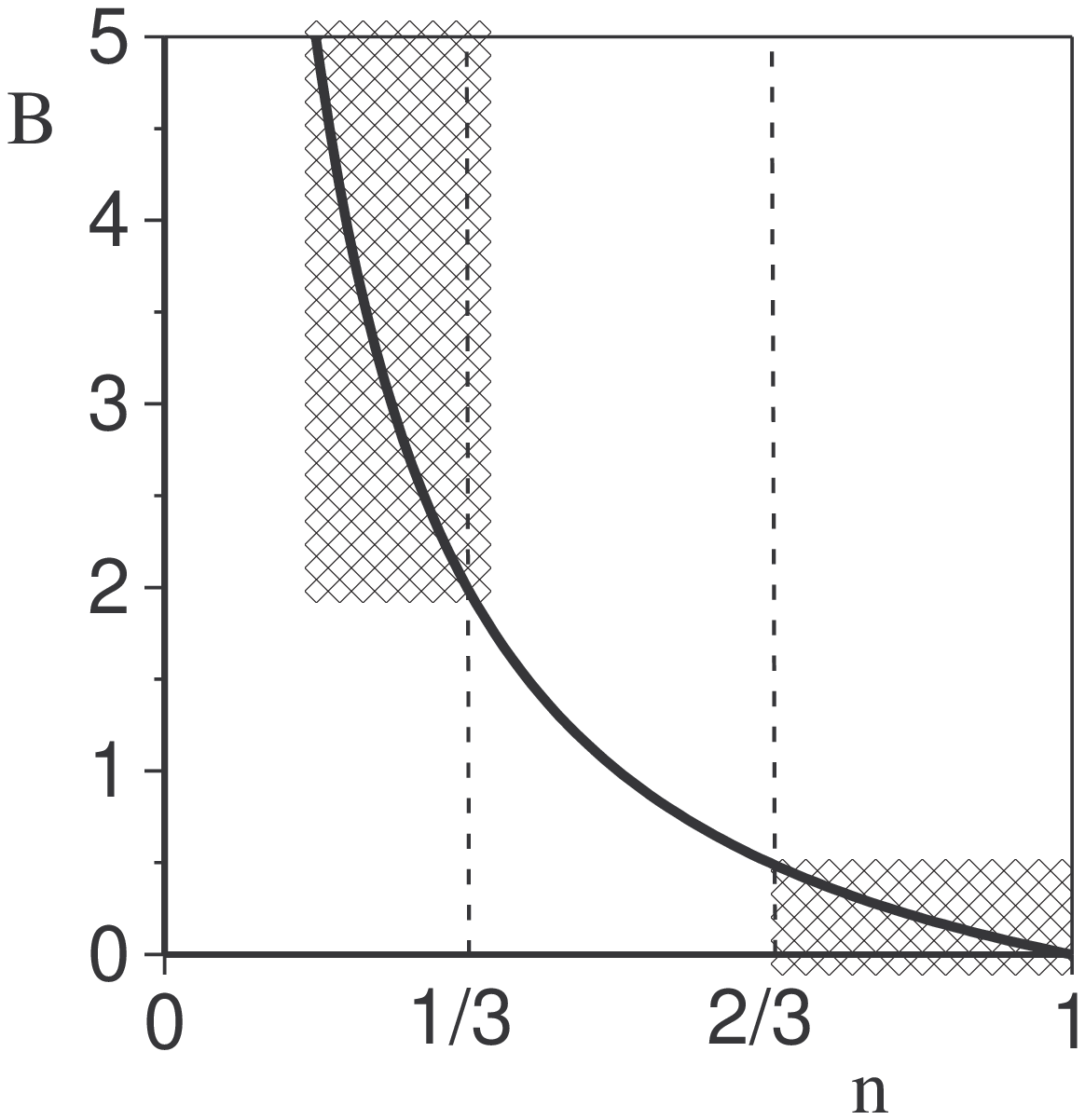}
\hfil
&
\hfil
\includegraphics[width=0.35\textwidth]{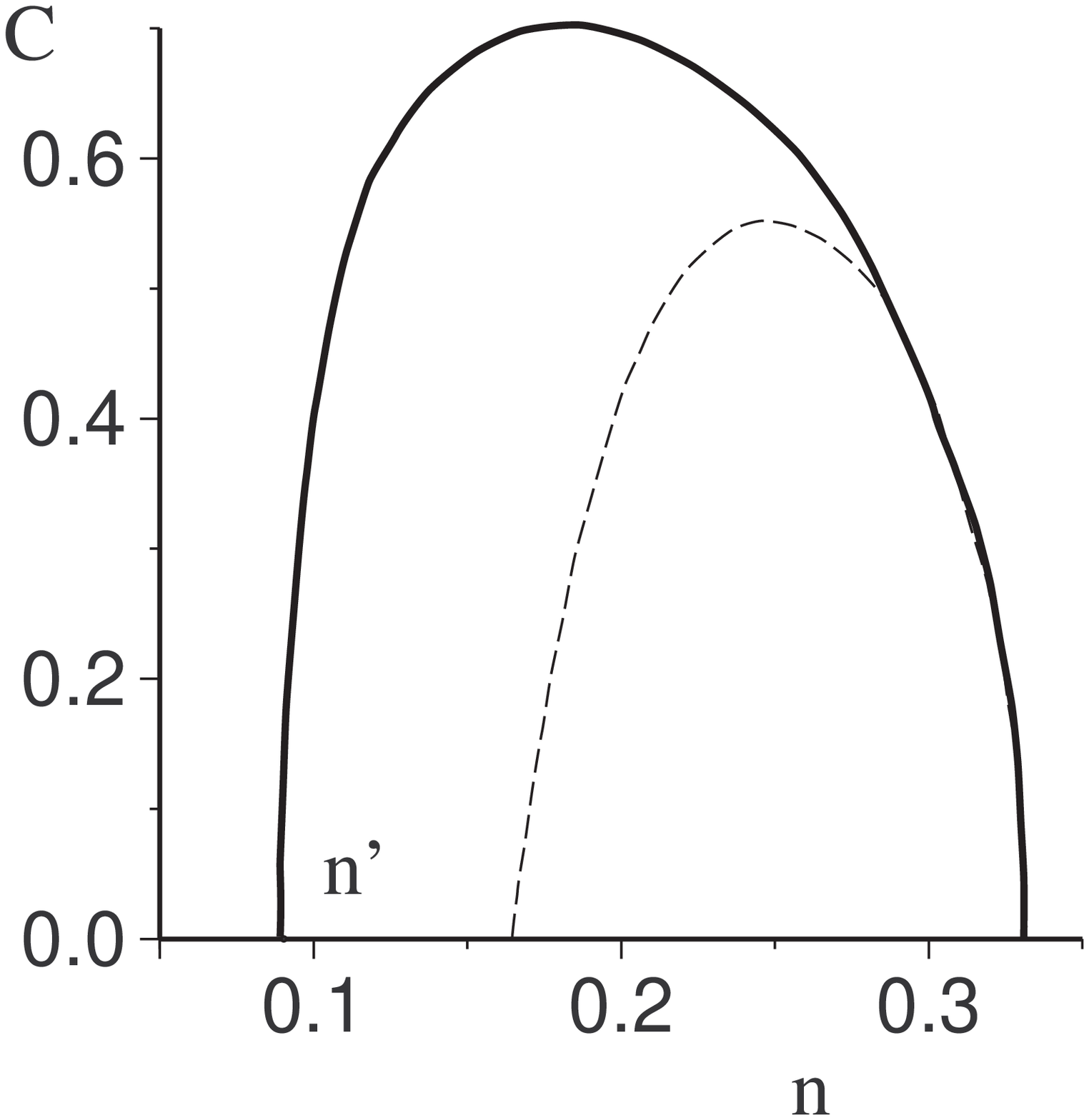}
\hfil
\\
\caption{%
Diagram of order of the paraelectric-ferroelectric phase transition at $C=0$.
Filled regions correspond to PT1, blank regions -- PT2.
}
\label{fig4}
&
\caption{%
Boundaries of PT1 region (inside the curves) at $B=5$. Solid line
corresponds to numerical calculations, dashed one -- the result of Landau
expansion analysis.
}
\label{fig5}
\end{tabular}
\end{center}
\end{figure}

It is convenient to make a qualitative analysis of PT2 region in the case
$C=0$, when the free energy (\ref{2.6}) has a diagonal form and OP is
equal to $\xi$ or $\eta$. The equation $B=1/n-1$ describes the curve, which
separates regions with different OP on the $B-n$ diagram. This equation can
be obtained by setting equal the temperatures at which coefficients at $\xi$
and $\eta$ in the quadratic form (\ref{2.6}) are equal to zero. It is found
that the sign of the coefficient at the fourth order of OP depends on $n$
only. Obtained diagram of a PT order is depicted in Fig.~\ref{fig4}.

\begin{figure}
\begin{center}
\begin{tabular}{p{0.45\textwidth}p{0.45\textwidth}}
\hfil
\includegraphics[width=0.4\textwidth]{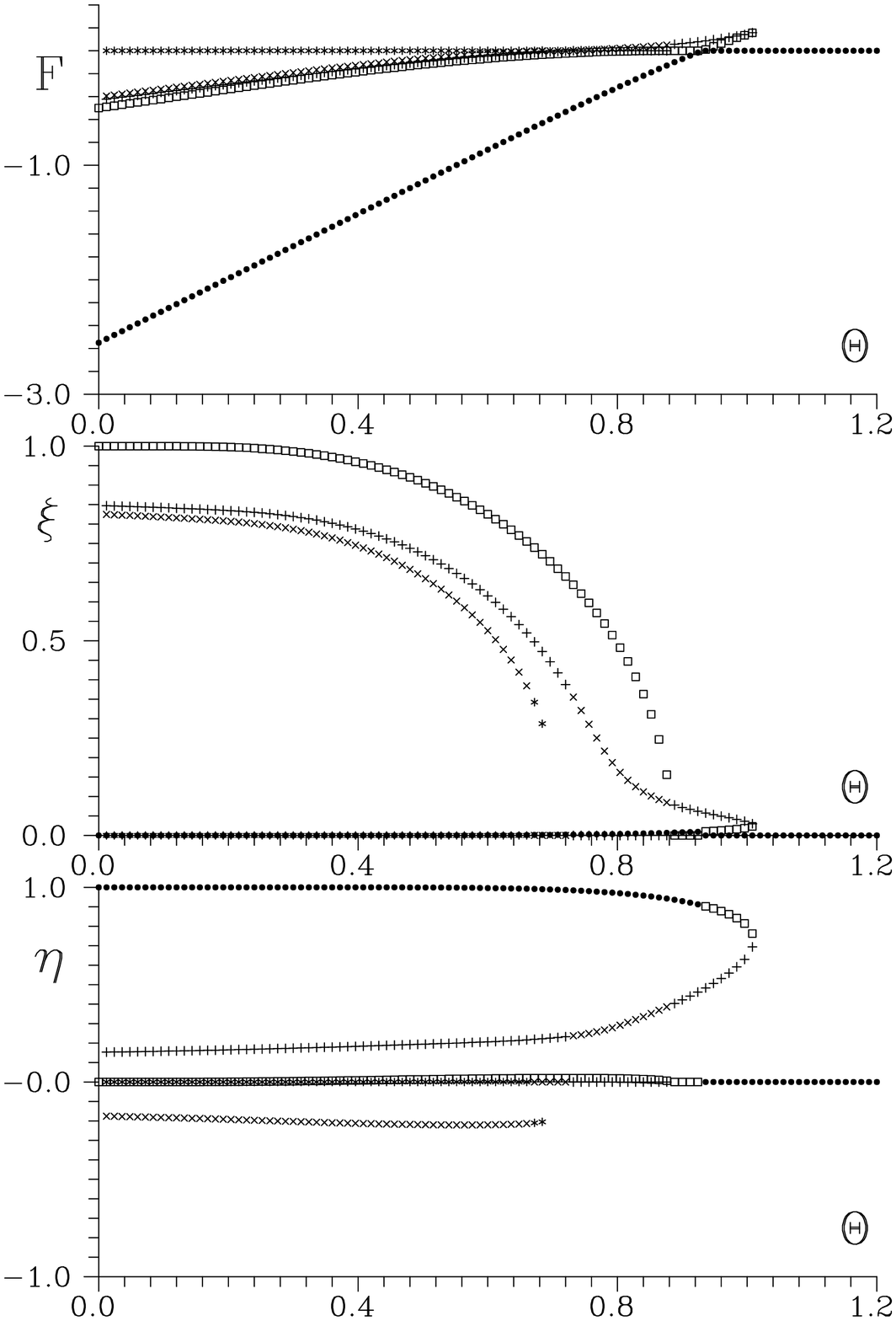}
\hfil
&
\hfil
\includegraphics[width=0.4\textwidth]{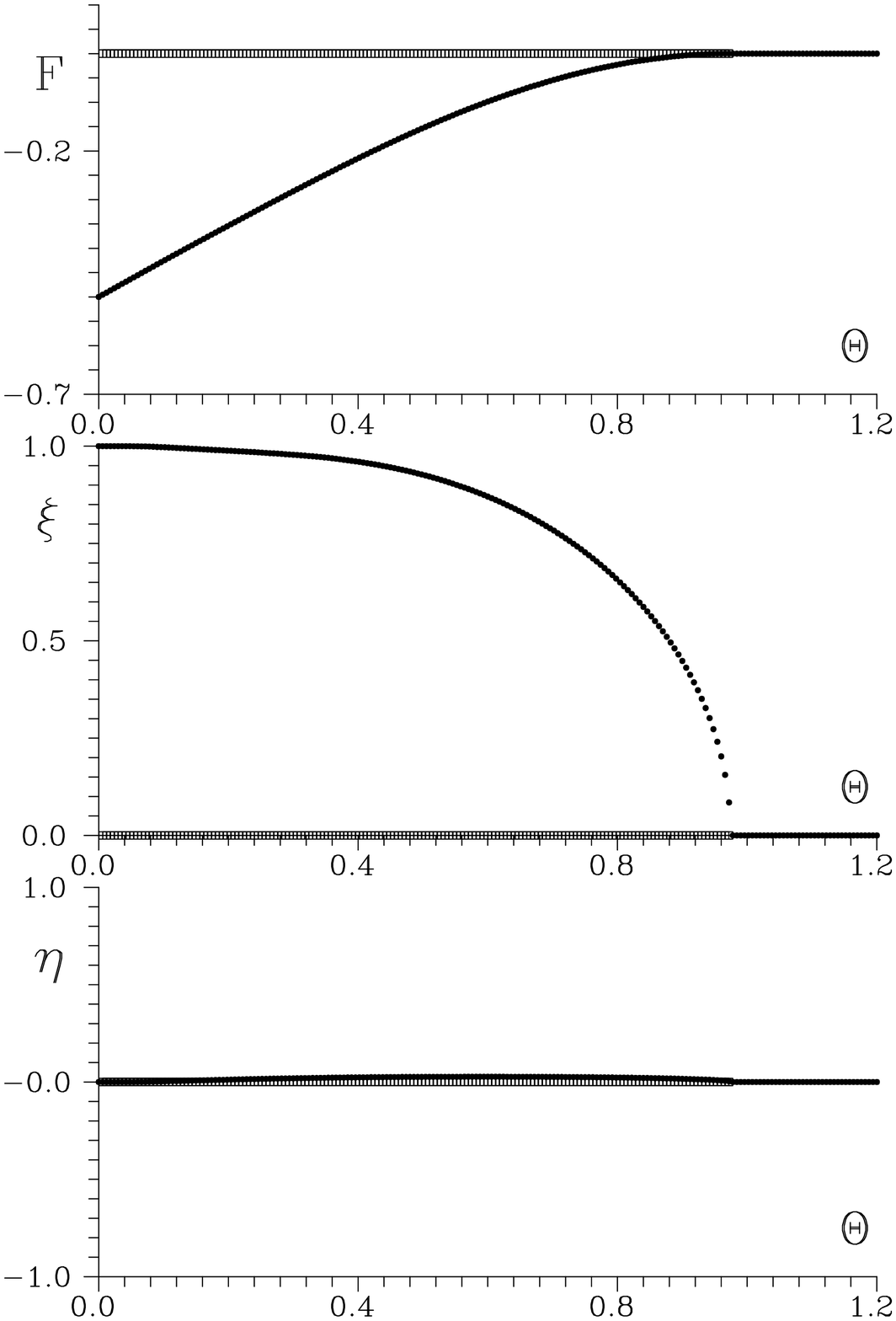}
\hfil
\\
\caption{%
Dependences of free energy and order parameters on temperature at values of
system parameters as
$B=5$, $C=0.1$, $n=0.12$.
}
\label{fig6}
&
\caption{%
\textit{Ibid}.
$B=0.5$, $C=0.7$, $n=0.05$.
}
\label{fig7}
\end{tabular}
\end{center}
\vspace{3ex}

\centerline{\includegraphics[width=0.4\textwidth]{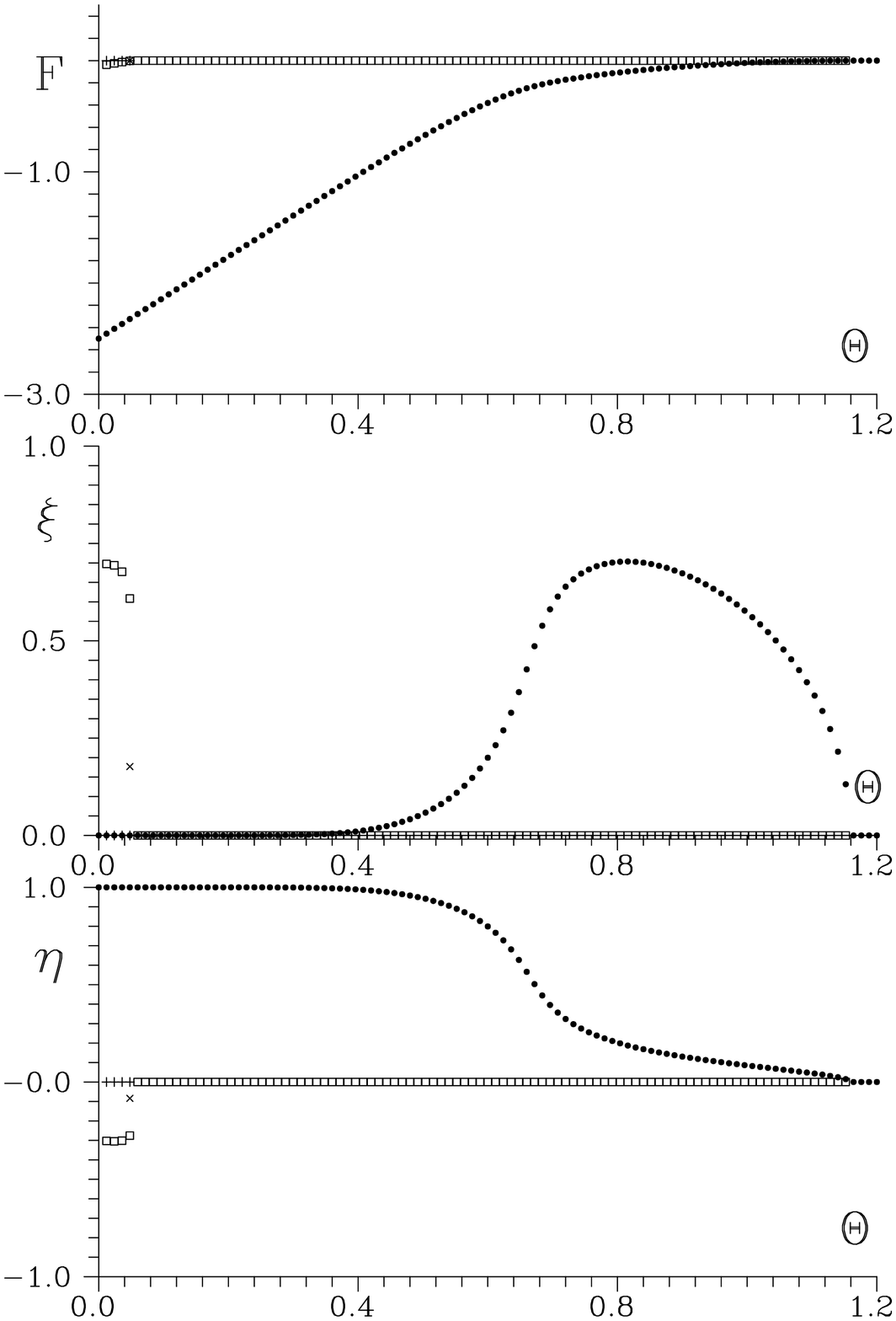}}
\caption{%
\textit{Ibid}.
$B=5$, $C=2$, $n=0.05$
}
\label{fig8}
\end{figure}

Increase of $C$ leads to the narrowing of the region of n where PT1 is
possible (Fig.~\ref{fig5}). One can see that the region, determined by the
$F_1^{(4)}$ sign change, is smaller than obtained by direct numerical
calculation.  This difference can be explained on the example of  the case
$C=0$. The point $n'$ is situated on the OP $\xi\to\eta$ change curve  of
the diagram in Fig.~\ref{fig4}. On the left from the $n'$ point the OP is
 $\xi$, hence PT2 on $\xi$ should takes place before PT2 in $\eta$.
But actually PT1 in $\eta$ takes place before them. It is illustrated in
Fig.~\ref{fig6} where dependences of free energy and parameters $\xi,\eta$
on temperature are depicted. It is also possible to find PT2 in $\xi$
firstly and PT1 with jump of $\eta$ at lower temperature. One should note
that at $C\neq{}0$ the OP $r_1$ is a mixture of $\xi$ and $\eta$. But at
considered here values of parameters the main contribution to the jump is
given by $\eta$. The diagram in Fig.~\ref{fig4} shows that PT1 with
prevailing jump of $\xi$ is typical for the region of small values of $B$
and large $n$; with prevailing jump of $\eta$ -- for large $B$ and small
$n$. Increase of $C$ cause a stronger ``mixing'' of $\xi$ and $\eta$, but
there is the limit $C^2<B$ (\ref{2.13}) of $C$ value. Thus it seems
difficult to find in this approach PT1 with the prevailing jump of $\xi$
(i.e. with the prevailing occupancies of positions 1 and 2) at small values
of $n$ what is characteristic of  DMAGaS crystals at PT from
paraelectric into ferroelectric phase. At small values of $n$ the model
exhibits PT2 (close to PT1) with prevailing change of $\xi$ parameter
(Fig.~\ref{fig7}).
So, we see that despite the simplicity of the model its thermodynamics is
very complicated. As an example an exotic case with sophisticated
dependencies of $\xi$ and $\eta$ on temperature is shown in Fig.~\ref{fig8}.

\section{Conclusions}

A microscopic approach based on the allowance for  different orientational states of
DMA ionic groups in crystal lattice is proposed in the present work for
description of phase transition in DMAAlS and DMAGaS crystals. In the
framework of a simple four-state model order parameters are introduced and
calculations of thermodynamical characteristics are made. The phase
transition from paraelectric to ferroelectric state is studied at different
relations between parameters describing interactions of orientational states of
different ionic groups.

An  interesting feature which follows from the symmetry analysis
is the strict correlation between the ordering along the
1--2~position axis (ferroelectric one) and the 3--4~position axis. Namely,
as follows from table~\ref{tab1}, ferroelectric ordering along the 1--2~axis
is accompanied by the antiferroelectric one along the 3--4~axis and vice
versa.

Ferroelectric state in these crystals is simultaneously improper
antiferroelectric one along an other axis and antiferroelectric state is
accompanied by ferroelectric ordering along the mentioned axis. Factors
affecting order of the phase transition at $T_{c1}$ are studied in more
details on the base of the model. At small values of $n$ this phase
transition is established to be of the second order close to the first order
one
(close to the tricritical point) under approximation that total occupancies
of positions (1,2) and (3,4) are assumed to be the same as in disordered
phase and temperature independent. This behaviour coincides with the known
result for DMAAlS crystal but does not match with data observed in DMAGaS
crystals. For the last case it should be necessary abandon the approximation
made and explicitly take into account temperature dependences of $n_{1}$
and $n_{2}$ values with the use of expressions~(\ref{add1}). Such an approach should
also be adopted for description of the low-temperature transition (at
$T_{c2}$) to antiferroelectric phase and investigation  of conditions of its
realization. This can be the closest perspective of development of the model.

\section{Acknowledgements}

The authors would like to thank Prof.~Z.Czapla and
Dr.~R.Tchuk\-win\-s\-kyi for their interest in the work and useful
discussions

This work was supported in part by the Foundation for
Fundamental Investigations of Ukrainian Ministry in Affairs of Science and
Technology, project No.~2.4/171.


\begin{thebibliography}{20}

\bibitem{PJCh93} Pietraszko~A., \L{}ukaszewicz~K., Kirpichnikova~L.F.~//
Polish J.\ Chem., 1993, vol.~67, p.~\mbox{1877--1884}.

\bibitem{PJCh94} Pietraszko~A., \L{}ukaszewicz~K.~// Polish J.\ Chem.,
1994, vol.~68, p.~1239--1243.

\bibitem{PJCh95} Pietraszko~A., \L{}ukaszewicz~K., Kirpichnikova~L.F.~//
Polish J.\ Chem., 1995, vol.~69, p.~\mbox{922--930}.

\bibitem{ZNat98} Tchukvinskyi~R., Cach~R., Czapla~Z.~// Z.~Naturforsch.,
1998, vol.~53a, p.~105--111.

\bibitem{r5} Andreev~E.F., Varikash~V.M., Shuvalov~L.A. //
Izv. AN~SSSR, ser. fiz., 1999, vol.~53, p.~572--574 (in Russian).

\bibitem{r6} Sobiestinskas~P., Grigas~Y., Andreev~E.F., Varikash~E.M. //
Phase Transitions, 1992, vol.~40, p.~85.

\bibitem{r7} Kapustianik~V., Bublyk~M., Polovinko~I., Sveleba~S.,
Trybula~Z., Andreev~E. // Phase Transitions, 1994, vol.~49, p.~231--235.

\bibitem{r8} Dacko~S., Czapla~Z. //
Ferroelectrics, 1996, vol.~185, p.~143--146.

\bibitem{r9} Pykacz~H., Czapla~Z. //
Ferroelectrics Letters, 1997, vol.~22, p.~107--112.

\bibitem{r11} Tchukvinskyj~R., Cach~R., Czapla~Z. //
Z.~Naturforsch., 1998, vol.~53a, p.~105--111.

\bibitem{r12} Czapla~Z., Tchukvinskyj~R. //
Acta Phys. Polonica~A, 1998, vol.~93, p.~527--530.

\bibitem{a1} Podsiad\l{}a~D., Czapla~Z., Andrievsky~B., Myshchyshyn~O. //
Acta Phys. Polonica~A, 1999, vol.~96, p.~409--416.

\bibitem{r10} Furtak~J., Czapla~Z. //
Ferroelectrics Letters, 1997, vol.~23, p.~63--67.

\bibitem{r14} Dolin\v{s}ek~J., Klanj\v{s}ek~M., Ar\v{c}on~D., Hae Jin Kim,
Selinger~J., \v{Z}agar~V. // Phys. Rev.~B, 1999, vol.~59, p.~3460--3467.

\bibitem{a2} Alsabbagh~N., Michel~D., Furtak~J., Czapla~Z. //
Phys. stat sol. (a), 1998, vol.~167, p.~77--87.

\bibitem{a3} Bednarski~W., Waplak~S., Kirpichnikova~L.F., Shuvalov~L.A. //
Phys. stat sol. (a), 1997, vol.~160, p.~R1--R2.

\bibitem{a4} Torgashev~V.I., Yuzyuk~Yu.I., Kirpichnikova~L.F.,
Shuvalov~L.A., Andreev~E.F. // Kristallografia, 1991, vol.~36, p.~677--685
(in Russian).

\bibitem{r16} Yasuda~N., Kaneda~A., Czapla~Z. //
Ferroelectrics, 1999, vol.~223, p.~71.

\bibitem{a5} Yasuda~N., Tajima~H., Czapla~Z. //
Physics Letters A, 1994, vol.~192, p.~137--140.

\bibitem{a6} Yasuda~N., Tajima~H., Czapla~Z. //
J.~Korean Phys. Soc., 1998, vol.~32, p.~S283--S285.

\bibitem{r13} Kazimirov~V.Yu., Rieder~E.E., Sarin~V.A., Belushkin~A.V.,
Shuvalov~L.A., Fykin~L.E., Ritter~C. // J.~Korean Phys. Soc., 1998,
vol.~32, p.~S91--S93.

\end{thebibliography}
\end{document}